\documentclass[aps,twocolumn,showkeys,preprintnumbers,superscriptaddress]{revtex4}

\usepackage{graphicx}		
\usepackage{dcolumn}		
\usepackage{bm}				
\usepackage{amsmath}
\usepackage{amssymb}
\usepackage{amsfonts}
\usepackage{amssymb}

\begin{document}

\title{Frequency shift keying in vortex-based spin torque oscillators}

\author{M. Manfrini}
\email{manfrini@imec.be}
\affiliation{IMEC, Kapeldreef 75, B-3001 Leuven, Belgium}
\affiliation{Laboratorium voor Vaste-Stoffysica en Magnetisme, K.U. Leuven, 3001 Leuven, Belgium}

\author{T. Devolder}
\author{Joo-Von Kim}
\author{P. Crozat}
\author{C. Chappert}
\affiliation{Institut d'Electronique Fondamentale, CNRS, UMR 8622, Orsay, France}
\affiliation{Univ. Paris-Sud, Orsay, France}

\author{W. Van Roy}
\affiliation{IMEC, Kapeldreef 75, B-3001 Leuven, Belgium}
 
\author{L. Lagae}
\affiliation{IMEC, Kapeldreef 75, B-3001 Leuven, Belgium}
\affiliation{Laboratorium voor Vaste-Stoffysica en Magnetisme, K.U. Leuven, 3001 Leuven, Belgium}

\date{\today}

\begin{abstract}
Vortex-based spin-torque oscillators can be made from extended spin valves connected to an electrical nanocontact. We study the implementation of frequency shift keying modulation in these oscillators. Upon a square modulation of the current in the $10\;\textrm{MHz}$ range, the vortex frequency follows the current command, with easy identification of the two swapping frequencies in the spectral measurements. The frequency distribution of the output power can be accounted for by convolution transformations of the dc current vortex waveform, and the current modulation. Modeling indicates that the frequency transitions are phase coherent and last less than 25 ns. Complementing the multi-octave tunability and first-class agility, the capability of frequency shift keying modulation is an additional milestone for the implementation of vortex-based oscillators in RF circuits.
\end{abstract}


\maketitle

\section{\label{sec:level1}Introduction}


In confined magnetic layers, the spontaneous generation of a vortex phase requires a proper ratio of lateral dimension and thickness. The stabilization of the vortex phase have been previously observed~\cite{Cowburn1999} and analytically investigated~\cite{Metlov2002,Guslienko2001,Hoffmann2002}. The vortex micromagnetic structure is very stable and consists of a magnetization that curls in the film plane around a central core magnetized out of the plane. This curling configuration avoids stray fields except at the core, and prevents the formation of domain walls~\cite{Shinjo2000,Wachowiak2002}. Early studies thus focused on the switching of the vortex core polarity, which was seen as a propitious candidate for non-volatile memory applications ~\cite{Yamada2007}. More recently, the dynamics of the vortex core has been studied while prospecting a new class of high frequency oscillators. Indeed vortex oscillation can be induced by dc-currents via the spin-transfer torque effect~\cite{Slonczewski1996,Berger1996}, an advantageous mechanism which does not require magnetic field and is scalable down to nanosystems.

Amongst spin-torque oscillators (STOs) that make use of vortex states, the microwave signatures emanating from the vortex motion have been experimentally studied in spin-valve nanopillars~\cite{Pribiag2007} and electrical nanocontacts systems in both frequency~\cite{Pufall2007,Mistral2008,Ruotolo2009} and time domains~\cite{Keller2009,DevolderAPL2009}. In the nanocontact geometry, the vortex dynamics have been formalized using the rigid vortex model~\cite{Mistral2008}. The formation of a confining potential created by the current ($I_{dc}$) permits the nucleation of a magnetic vortex~\cite{Guslienko2002} that is spin-transfer torque driven to a large orbit around the nanocontact in the low-frequency interval ($100-600\;\textrm{MHz}$). The relative change in the magnetization direction of the free layer with respect to the pinned layer is translated to a time-varying voltage via the giant magnetoresistance (GMR). Currently, the power transducing yield of an oscillating vortex in the nanocontact configuration is greater than that of uniform ferromagnetic modes, due to the nearly complete rotation of the magnetization translated by GMR.~\cite{Lehndorff2009}.

A pivotal result from both theory and experiment~\cite{Mistral2008} is the quasi-linear frequency tunability with the current. The present theoretical understanding is that an almost fixed trajectory is maintained while the vortex gyrates under distinct currents. We have studied this orbital stability in a previous work ~\cite{Manfrini2009}, employing a microwave interferometer to insert MHz-modulated currents ($I_{mod}$) into the device and hence study the resulting vortex dynamics. The agility, i.e., the shortest time it takes for the vortex to hop and stabilize from one frequency to another was found to be below a maximum upper bound value of $20\;\textrm{ns}$, comparable to the agility of state of the art voltage-controlled oscillators~\cite{Grebennikov2007}. Studies on the ability of modulating nanocontact STOs are necessary assessments for a more compatible integration with the current silicon complementary metal-oxide semiconductor industry. For instance, by rapidly swapping the magnetic vortex motion between two well-defined frequencies is a major asset for developments on telecommunication devices operating at the low frequency regime.

Here, we investigate the implementation of a frequency shift keying (FSK) modulation scheme of a vortex-based nanocontact oscillator over a large current range. Upon the MHz-current modulation the vortex experiences two distinct current states yielding largely separated frequencies. For currents modulated at low pulse repetition frequencies (PRF), the vortex gyration frequency follows the current command, with easy identification of the two swapping frequencies in the spectral measurements. For currents modulated at higher PRF, the modes in the power density spectrum split and multiple sidebands appear. The frequency distribution of the output power can be accounted for analytically by convolution transformations of the dc current vortex waveform, and the current modulation. This indicates how abrupt frequency transitions can be considered.

\section{\label{sec:level2}Device and Experimental Setup}
Our devices are electrical nanocontacts with a physical radius of $60\;\textrm{nm}$, established to the top of a bottom-pinned exchange-biased spin-valve. The multi-layered stack is dc-sputtered in the following composition (thickness in nanometers): [Ta (3.5)/Cu (16)]$_{\times2}$/Ta (3.5)/Ni$_{80}$Fe$_{20}$ (2) as bottom electrode. The spin-valve is Ir$_{22}$Mn$_{78}$ (6)/Co$_{90}$Fe$_{10}$ (4.5)/Cu (3.5)/Co$_{90}$Fe$_{10}$ (1.5)/Ni$_{80}$Fe$_{20}$ (2), and it is capped with Ta (1.5)/Pt (4) to prevent oxidation. Device fabrication, magnetic, and electrical properties can be found elsewhere~\cite{DevolderSPIE2009}. The series resistance is $R=9\;\Omega$ and the GMR value is $\Delta R=25\;\textrm{m}\Omega$. Microwave measurements have been performed in zero applied magnetic field. The electrical current is applied perpendicular to the film plane, and positive current is defined as electrons flowing from the free to the pinned magnetic layer. Current-induced vortex oscillations are observed after nucleation at $I_{dc}= 50\;\textrm{mA}$ and the vortex magnetization dynamics can be followed until when $I_{dc}$ drops below $9\;\textrm{mA}$, consistent with previous findings~\cite{DevolderAPL2009}.\\
Once the magnetic vortex motion is established, the current gets modulated through a proper experimental set-up~\cite{Manfrini2009} ensuring both satisfactory impedance matching and cancellation of the modulating signal at the front end of the spectrum analyzer. The raw data spectra are converted to power density spectra by assuming a frequency-independent noise figure of the amplifiers, and an imperfect but reproducible cancellation of the modulating signal.
The peak-to-peak modulation corresponds to $2I_{mod} \approx 10.0\;\textrm{mA}$. We observe how this square pulse modulation of the current affects the dynamics of the vortex, for various pulse repetition frequency (PRF) ranging from 0 to 30 MHz. Therefore, every first half of the pulse period $t_{pulse}=1/\textrm{PRF}$, the vortex is subjected to $I_{dc}-I_{mod}$ and orbits with gyration frequency $F_1$. Every second half of the pulse period, the current is $I_{dc}+I_{mod}$ and leads to a vortex gyration frequency $F_2$. The two applied currents are separated by transitions with 2 ns rise (or fall) times.

\begin{figure}
\includegraphics{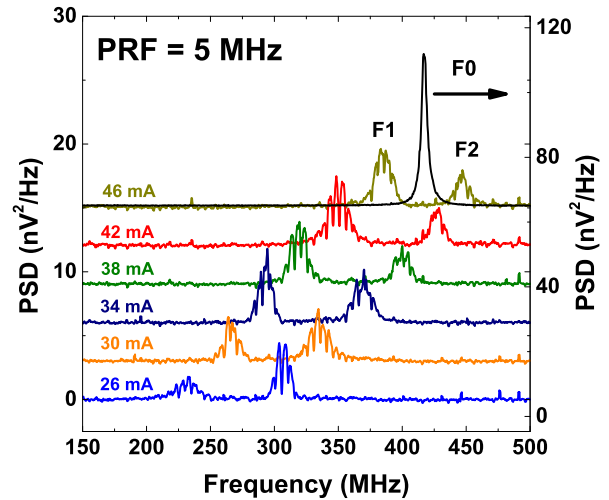}
\caption{(Color online) Power spectral densities for several values of $I_{dc}$ in constant-current operation (bold $F_0$ line and right axis) or in modulated conditions. For a PRF of $5\;\textrm{MHz}$, the electrical signature of the vortex motion swaps from $F_{1}$ to $F_{2}$. Increasing values of current lead to a linear blue-shift for $F_{1}$ and $F_{2}$. The curves were smoothed with an aperture of $11\;\textrm{kHz}$ and have been shifted vertically by $+3\;\textrm{nV\textsuperscript{2}/Hz}$ from each other. 
\label{fig:one}}
\end{figure}

\section{\label{sec:level3}Experimental Results}
\begin{figure}
\includegraphics{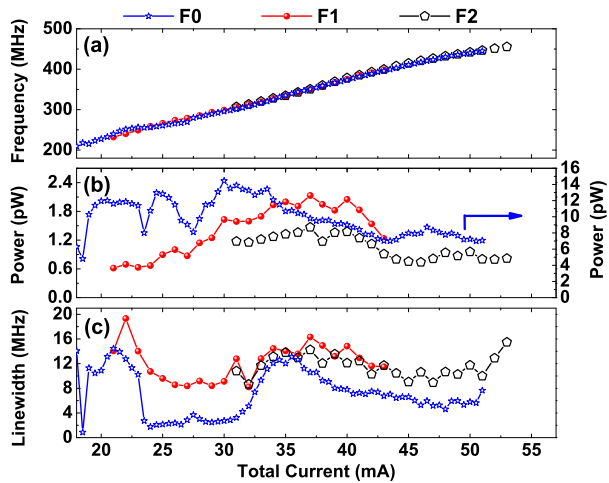}
\caption{(Color online) Frequency~\textbf{(a)}, power~\textbf{(b)} and linewidth~\textbf{(c)} of a vortex oscillator subjected to PRF of $5\;\textrm{MHz}$. The curves for $F_1$ have been displayed as function of $I_{dc}+I_{mod}$, whilst the curves for $F_2$ have been generated with $I_{dc}-I_{mod}$.}
\label{fig:two}
\end{figure}

\begin{figure}
\includegraphics{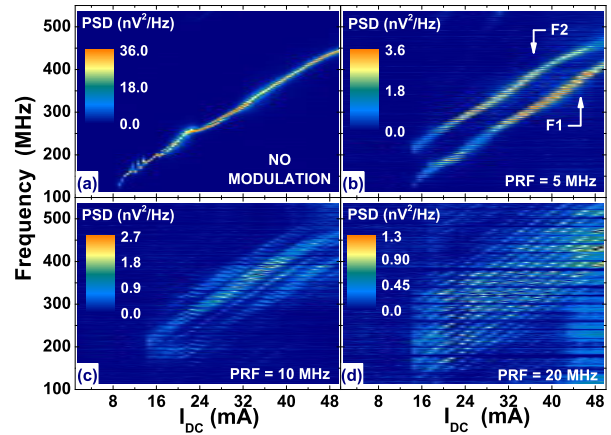}
\caption{(Color online) Current evolution of the power spectral density. In the absence of current-modulation~\textbf{(a)} the vortex oscillates in a quasi-linear frequency fashion with frequency $F_0$. Under the influence of a small PRF value~\textbf{(b)}, such as $5\;\textrm{MHz}$, the main mode splits in $F_1$ and $F_2$ branches that can be clearly resolved in frequency with an average peak-to-peak separation of $72\;\textrm{MHz}$. For a high PRF value as $20\;\textrm{MHz}$ in~\textbf{(d)} the frequency separation is no longer clear and the spectra becomes saturated with multiplets from generated sidebands. The transition for the frequency resolution occurs for a PRF of $10\;\textrm{MHz}$ in~\textbf{(c)}. 
\label{fig:three}}
\end{figure}

Fig.~\ref{fig:one} depicts the power spectral density (PSD) of the modulated vortex oscillator for several values of $I_{dc}$ and a fixed PRF of $5\;\textrm{MHz}$.
Fig.~\ref{fig:two} reports the figures of merit (frequency, power and linewidth) of each mode identified from spectra similar to these of Fig.~\ref{fig:one}. Fig.~\ref{fig:three} finally displays how the PRF influences the spectral distribution of the output power of the oscillator.

In Fig.~\ref{fig:one}, the top curve is the spectrum of a vortex oscillating with a (single) frequency $F_0$ at a fixed (unmodulated) current of $46\;\textrm{mA}$. Once the modulation is switched on for a PRF of $5\;\textrm{MHz}$, the nanocontact is subjected to a current $I_1=I_{dc}- I_{mod}$ during $100\;\textrm{ns}$ and a current $I_2= I_{dc}+ I_{mod}$ during the next $100\;\textrm{ns}$. The $F_0$ peak splits and the oscillator power gets distributed among two peaks at $F_1$ and $F_2 > F_1$, each of them carrying a shared, hence reduced power. The two peaks can be resolved, which indicates that the necessary time for the vortex frequency to stabilize in one of these frequencies (the agility) is much shorter than $100\;\textrm{ns}$. For large current values ($I_{dc}= 46\;\textrm{mA}$), the output power is mainly carried by $F_1$. As $I_{dc}$ decreases, the output power is being transferred from $F_1$ to $F_2$ until most of the power is carried by $F_2$ as shown at $I_{dc}= 26\;\textrm{mA}$. We will see below that this is provenient from the specificities of the dependence of power on current in the constant current mode.
The peak-to-peak separation ($F_2 - F_1$) is $72\;\textrm{MHz}$ on average, with some dependence on $I_{dc}$. The smallest peak-to-peak separation is $58\;\textrm{MHz}$ at high dc-current and it is correlated to a point of inflection in the frequency versus current curve in the constant current mode (Fig.~\ref{fig:three}a). \\ 
Let us define $\Delta F_1$ and $\Delta F_2$ the full width at half maximum linewidths of the envelopes of the modes at $F_1$ and $F_2$ depicted in Fig.~\ref{fig:one}. One can note that for a low value of PRF as $5\;\textrm{MHz}$, we have $F_2 - F_1\gg\Delta F_1\sim\Delta F_2\gg\textrm{PRF}$, leading to a clear distinction of the peaks $F_1$ and $F_2$ whatever the value of the dc current (Fig.\ref{fig:one}). 

Let us detail how the current modulation $I_{mod}$ affects the vortex dynamics. For this, we extract the figures of merit for the vortex oscillator (frequency, power and linewidth) from the spectra subjected to a PRF of $5\;\textrm{MHz}$ by using a double-Lorentzian fitting tool. To confirm our estimate of $I_{mod}$, we plot frequency (Fig.~\ref{fig:two}a), power (Fig.~\ref{fig:two}b) and linewidth (Fig.~\ref{fig:two}c) of $F_1$ and $F_2$ modes as function of $I_{dc}+I_{mod}$ and $I_{dc}-I_{mod}$, respectively. The criterion to determine $I_{mod}$ has been defined by the best overlapping with $F_0$ of the curves $F_1$ and $F_2$, as depicted in Fig.~\ref{fig:two}a. The $F_1$ and $F_2$ curves overlap satisfactorily. We observe a reduction in linewidth; this feature was already present without modulation (blue curves in Fig.~\ref{fig:two}a, b and c), and is most probably due to microstructural effects. On the other hand, the fact that the total power is shared between the two main frequencies when the modulation is on, leads to a more complex overlapping of the power curves, as in Fig.~\ref{fig:two}b. The overlapping indicates that during each instantaneous value of the applied current $I_1$ or $I_2$, the main signatures of the vortex dynamics (frequency, power and linewidth) are very similar to what they were in constant-current operation. 

The evolution of the voltage noise power spectra has been studied for PRF ranging from $0.5$ to $30\;\textrm{MHz}$ (Fig.~\ref{fig:three}). Panel~\ref{fig:three}a shows the current-dependence of the noise power densities for a free running vortex without current modulation. The tunability is $dF_{0}/dI_{dc}=7.6\;\textrm{MHz/mA}$ and the signature of the vortex gyration is detectable down to $9\;\textrm{mA}$. When a current modulation with PRF $ \leq 5\;\textrm{MHz}$ is applied, we observe two frequency branches $F_1$ and $F_2$ witnessing the vortex dynamics (Fig.~\ref{fig:three}b). Increasing the PRF leads to a richer spectra, and it is no longer that easy to identify the two different gyration modes at first sight.  The transition occurs for a PRF of $10\;\textrm{MHz}$ (Fig.~\ref{fig:three}c), when $\Delta F_1\sim\Delta F_2\sim\textrm{PRF}$. Sidebands are now well separated from the main peaks $F_1$ and $F_2$, and appear as multiplets.
An even stronger distortion is observed for higher values of PRF, at which almost the entire spectra get flooded with sidebands, as in panel~\ref{fig:three}d for a PRF of $20\;\textrm{MHz}$. Remarkably, now the current is changed as rapidly as every $25\;\textrm{ns}$, which corresponds to merely ten full gyrations of the vortex around the nanocontact. We could resolve sidebands until a PRF of $30\;\textrm{MHz}$ (not shown). Larger values of PRF lead to experimental artifacts that dominate over the signal of interest. Note finally that when modulating the current, the vortex modes disappear at $I_{dc}-I_{mod}=10.5\;\textrm{mA}$. This value is slightly greater than the annihilation current observed in the absence of modulation ($9\;\textrm{mA}$). The reason for this difference is not understood. 

\section{\label{sec:level4}Discussion}
The discussion of the experimental spectra requires a brief analysis of the expected spectrum of a frequency shift keyed vortex oscillator. In the ideal case, the frequency transition would be infinitely abrupt at the current changes and it would be continuous in phase. In practice, we observe the transition being strongly dependent on how fast the vortex is modulated, i.e., a relation between the PRF and linewidths $\Delta F_1$ and $\Delta F_2$.

\subsection{\label{sec:phaselost}Discontinuous-phase frequency shift keying}

When $\textrm{PRF}\gg\Delta F_1$, $\Delta F_2$, the phase memory is not preserved between each of the oscillating half-period. Thus, the time-resolved voltage $v_{\textrm{FSK}}(t)$ could be written in the form:

\begin{equation}
v_{\textrm{FSK}}(t) =  \sqcap(t) [v_{F1}(t) - v_{F2}(t)] + v_{F2}(t)\;\;,
\label{vortex_signal}
\end{equation}

where $v_{F1}(t)$ and $v_{F2}(t)$ are the voltage waves emitted by the vortex in dc currents $I_1$ and $I_2$, i.e., approximately sine waves~\cite{DevolderAPL2009} at frequencies $F_1$ and $F_2$ subjected to phase noise, and where $\sqcap(t)$ is a square wave swapping between 0 and 1 at a frequency PRF.
The complex spectrum of the voltage $V_{\textrm{FSK}}(f)$ is the convolution product given by:

\begin{equation}
V_{\textrm{FSK}}(f) = \sqcap(f) \otimes [V_{F1}(f) - V_{F2}(f)] + V_{F2}(f)\;\;,
\label{convolution}
\end{equation}

where the frequency comb generated by the square wave is~\cite{Arfken2005}:

\begin{equation}
SQ(f)=\frac{1}{2} + \frac{1}{\pi} \sum^{+ \infty}_{\underset{n = \textrm{odd}}{n = - \infty}} \frac{(-1)^\frac{n-1}{2}}{n} \delta_{n \textrm{PRF}}\;\;.
\label{square_wave}
\end{equation}

The complex spectrum of the voltage $V_{\textrm{FSK}}(f)$ can be rewritten in the form:

\begin{eqnarray}
V_{\textrm{FSK}}(f) &=& \frac{V_{F1}(f)+ V_{F2}(f)}{2} + \nonumber \\
&+&\Bigg[\frac{1}{\pi} \sum^{+ \infty}_{\underset{n = \textrm{odd}}{n = - \infty}} \frac{(-1)^\frac{n-1}{2}}{n} V_{F1}(f-n\textrm{PRF})\Bigg] - \nonumber \\
&-&\Bigg[\frac{1}{\pi} \sum^{+ \infty}_{\underset{n = \textrm{odd}}{n = - \infty}} \frac{(-1)^\frac{n-1}{2}}{n} V_{F2}(f-n\textrm{PRF})\Bigg] \label{eq:complex_spectrum}.
\end{eqnarray}

The translation from complex spectra (Eq.\ref{eq:complex_spectrum}) to noise power densities simplifies only when there is no overlap between all spectral lines in the sum. This requires $F_2 - F_1\gg\textrm{PRF} > \Delta F_1$, $\Delta F_2$. These two conditions are satisfied for a PRF of 5 MHz,(Fig.~\ref{fig:three}b). In such case, each power spectral density should consist of the sum of two individually-symmetric sets of sidebands (multiplets). Each multiplet has the following features:
\begin{itemize}
\item The multiplets are essentially triplets centered around $F_1$ and $F_2$, with sidebands at $F_1 - \textrm{PRF}$, $F_1$ and $F_1 + \textrm{PRF}$ (idem for $F_2$), with intensities $1/\pi^2$, $1/4$ and $1/\pi^2$;
\item There are no sidebands at $F_1 - 2\textrm{PRF}$ and $F_1 + 2\textrm{PRF}$, as well as, for every even number of $n$ (idem for $F_2$). 
\item The next sideband peaks are at $F_1 \pm n\textrm{PRF}$ (idem for $F_2$) with $\textrm{n}={3,5,7,...}$, but they carry such a small power of $1/(n\pi)^2$ that they should hardly be observable.
\end{itemize}

\begin{figure}
\includegraphics{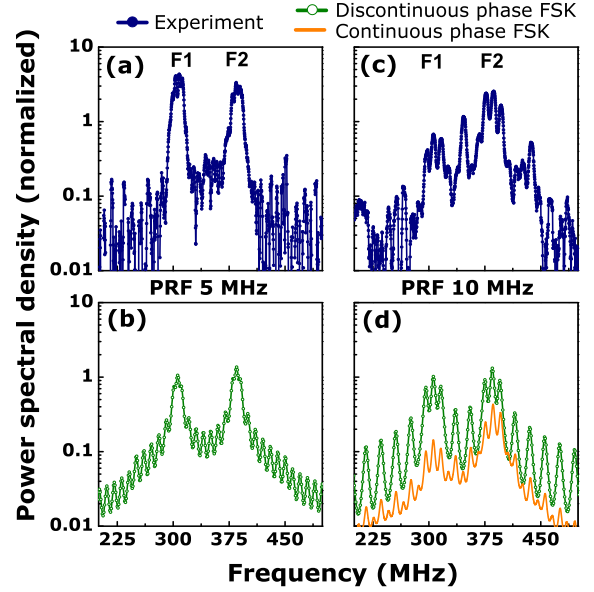}
\caption{(Color online) Experimental (top panels) versus theoretically expected (bottom panels) power spectra for pulse repetition frequencies of $5$~\textbf{(a, b)} and $10$~\textbf{(c, d)}. The solid green curves with open circles are from Eq.~\ref{eq:complex_spectrum}, i.e., assuming instantaneous frequency transitions and immediate stabilization. The solid orange curve is from Eq.~\ref{eq:complex_spectrum_cont} which preserves phase between frequency transitions.}
\label{fig:four}
\end{figure}

These expectations can be compared with experimental spectra (Fig.~\ref{fig:four}, top panels). We have performed the numerical evaluation of Eq.~\ref{eq:complex_spectrum} for parameters being those of $I_{dc}=36\;\textrm{mA}$ and $I_{mod}=5\;\textrm{mA}$. We thus assume Lorentzian lines for $V_{F1}(f)^2$ and $V_{F2}(f)^2$, with respective frequencies and linewidths being $F_1=306\;\textrm{MHz}$, $F_2=385\;\textrm{MHz}$, $\Delta F_1\approx \Delta F_2=10\;\textrm{MHz}$. Experimental (filled circles) and theoretical spectra (open circles) match nicely for a PRF of $5\;\textrm{MHz}$. This indicates that the oscillator frequency reaches a stabilized value during the plateaus of the current, i.e., in much less than $100\;\textrm{ns}$. However when the PRF is increased to $10\;\textrm{MHz}$ the experimental spectrum starts to exhibit a peak in between the two triplets, and this peak is not accounted for by Eq.~\ref{eq:complex_spectrum}. We will see that this peak arises from the fact that when the PRF is large enough, the phase is essentially preserved during the modulation period $1/PRF$. This leads us to an upgraded description for $PRF \geq 10\;\textrm{MHz}$.

\subsection{\label{sec:phasecontinuous}Continuous-phase frequency shift keying}

Now the condition $\textrm{PRF}\sim\Delta F_1$, $\Delta F_2$ is achieved, such that the phase memory is similar or longer than the modulation period. Therefore, a model incorporating phase continuity needs to be developed. The time-integrated continuous-phase output signal~\cite{Lucky1968} of a vortex oscillator can be represented as

\begin{equation}
v_{\textrm{FSK}}(t) = \cos\Bigg[F_{0}t + \frac{F_2 - F_1}{2} \wedge(t)\Bigg] \;\;,
\label{eq:vortex_signal_cont}
\end{equation} 

where $F_0=\frac{F_1+F_2}{2}$ is the mean frequency and $\wedge(t)$ is a triangular function with period $\frac{1}{PRF}$ and ranges between $\pm\frac{1}{4PRF}$. Deriving the complex spectrum of the continuous-phase voltage $V_{\phi}(f)$ results in

\begin{equation}
V_{\phi}(f) = \sum_{n\in\mathbb{Z}}C_{n}\;\delta_{\omega_{0}+n\textrm{PRF}}(f)\;\;,
\label{eq:complex_spectrum_cont}
\end{equation}

with

\begin{equation}
C_{0} = \frac{8\textrm{PRF}}{F_2 - F_1}\sin\Bigg[\frac{F_2 - F_1}{8\textrm{PRF}}\Bigg]\;\;,
\label{eq:first_constant}
\end{equation}

given that the frequency spacing $F_2 - F_1\gg2\textrm{PRF}$. Note that equation~\ref{eq:complex_spectrum_cont} includes both odd and even values of $n$. Therefore, the former triplets around $F_1$ and $F_2$ transform in multiplets, yielding a vortex oscillating at a mean frequency $F_0$. The appearance of a central peak in the experimental spectrum at $346\;\textrm{MHz}$ (Fig.~\ref{fig:four}c) with the generation of sidebands around $F_0$ is therefore an indicator that phase is preserved during the frequency transitions, which is the main outcome of our results for the PRF of $10\;\textrm{MHz}$. Note that a suppression of the central peak can only occur if the condition $n(F_2-F_1)>4\textrm{PRF}$ for any $n$ non-zero integer is met, which is not the case discussed here.\\ 
Increasing the PRF to $15$ (not shown), $20$, and then $25\;\textrm{MHz}$ (Fig.~\ref{fig:five} a, c), the experimental spectra no longer match with the expectation from Eq.~\ref{eq:complex_spectrum} nor from Eq.~\ref{eq:complex_spectrum_cont}. This indicates that the assumption of abrupt frequency transitions and immediate frequency stabilization is no longer adequate at those timescales.

\subsection{\label{sec:fm} Frequency modulation in vortex oscillators}

To understand our experimental results at large PRF, let us recall the expected shape of the spectra in the opposite limit when the oscillator frequency does not succeed in stabilizing during $I_1$ and $I_2$, but oscillates around a mean frequency value $F_0=\frac{F_1+F_2}{2}$. An extreme case would be when the oscillator frequency lags behind the modulation and varies sinusoidally. The instantaneous frequency would then be $F(t)=\frac{F_0}{2} + \beta \textrm{PRF} \sin(2 \pi\textrm{PRF} t)$, where $\beta \leq \frac{F_2-F_1}{\textrm{PRF}}$ is the modulation index. The complex spectrum of such a frequency-modulated (FM) signal can be described as~\cite{Carson1922}:

\begin{equation}
V_{FM}(f)=\sum^{+ \infty}_{n = - \infty}  {{J_n(\beta)}}  V_{F0}(f - n\textrm{PRF})\;\;,
\label{eq:FM_spectrum}
\end{equation}

where ${J_n}$ is the $n^{th}$ Bessel function. It is worth comparing Eq.~\ref{eq:FM_spectrum} to Eq.~\ref{eq:complex_spectrum}, and noticing the following facts:

\begin{itemize}
\item In case the frequency variation is gradual (Eq.~\ref{eq:FM_spectrum}) instead of abrupt (Eq.~\ref{eq:complex_spectrum}), even and odd-numbered sidebands are \emph{both} present. Since the modulation index $\beta$ is large in our case (i.e. $\beta > > 1$), even and odd- numbered sidebands have similar amplitudes, globally decaying with $n$;
\item Also, power occupies a frequency window larger than in the pure FSK case: Carson's rule~\cite{Carson1922} indicates that 98\% of the power is in a band of $2(F_2-F_1)+2PRF$, in practice circa 200 MHz.
\end{itemize}

\begin{figure}
\includegraphics{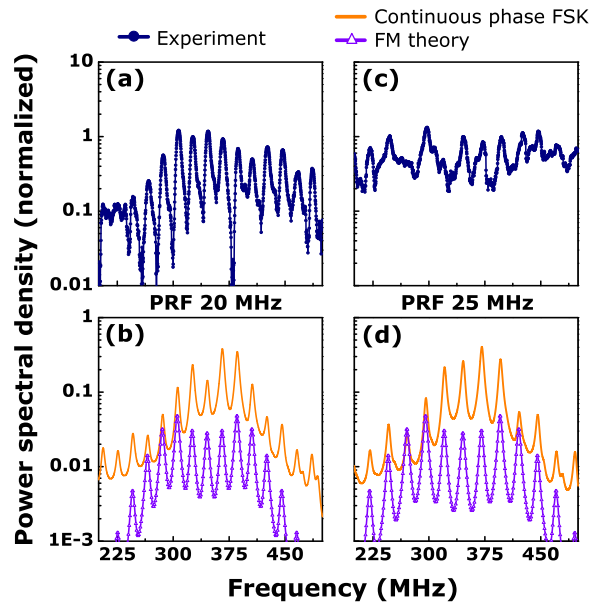}
\caption{(Color online) Experimental (top panels) versus theoretically expected (bottom panels) power spectra for pulse repetition frequencies of $20$~\textbf{(a, b)} and $25\;\textrm{MHz}$~\textbf{(c, d)}. The solid orange curves are continuous-phase spectra from Eq.~\ref{eq:complex_spectrum_cont}. An immediate frequency stabilization is not possible due to the high PRF value. Gradual frequency transitions occurs now, and a model of frequency modulation is required. The violet open triangle symbol curves are from Eq.~\ref{eq:FM_spectrum}.}
\label{fig:five}
\end{figure}

We have evaluated the spectrum expected for pure frequency modulation (Eq.~\ref{eq:FM_spectrum}) for a PRF equal to $20\;\textrm{MHz}$ and $25\;\textrm{MHz}$, using a large modulation index $\beta=19$. The behavior are displayed in Fig.~\ref{fig:five} b, d (open triangles for FM theory). The assumption of gradual frequency transition lasting $25\;\textrm{ns}$ and $20\;\textrm{ns}$ clearly matches with the experimental data much better than did Eq.~\ref{eq:complex_spectrum_cont}. This indicates that the time needed for frequency stabilization after an abrupt change of the current is $25 \pm 5\;\textrm{ns}$.
\section{\label{sec:level5} Conclusions}
In this work, we have studied vortex-based spin-torque oscillators made from extended spin valves connected to an electrical nanocontact. We have shown that Frequency Shift Keying modulation can be implemented in these oscillators, and that the identification of the two swapping frequencies can be done directly in the spectral domain. Upon a square modulation of the current in the $0-10\;\textrm{MHz}$ range, the vortex instantaneous frequency seems to follow the current command with a stabilization time of circa $25\;\textrm{ns}$, while the phase is preserved at the transition. Indeed at these modulation frequencies, the frequency distribution of the output power can be perfectly accounted for by convolution transformations of the voltage waveform of the gyrating vortex in dc current, and the current modulation, while at higher modulation frequencies, the oscillator frequency does not longer follow entirely the current. The possibility of implementing a frequency shift keying in magnetic vortex systems endows these oscillators as prospective candidates for in applications requiring compacity, tunability and frequency modulation based communication schemes.


\begin{acknowledgments}
The authors thank E. Vandenplas and J. Feyaerts for technical support, and J. Moonens for electron-beam lithography. M. M. is supported by the European Community under the 6\textsuperscript{th} FP for the Marie Curie RTN SPINSWITCH, contract n$^{\circ}$ MRTN-CT-2006-035327. This work has been supported by the Triangle de la Physique contract 2007-051T.
\end{acknowledgments}



\end{document}